\documentstyle[graphicx,amsmath,amsfonts]{article}
\def\be{\begin{equation}}
\def\ee{\end{equation}}
\def\ba{\begin{array}}
\def\ea{\end{array}}
\def\bea{\begin{eqnarray}}
\def\eea{\end{eqnarray}}
\def\d{{\rm d}}
\def\e{{\rm e}}
\def\exp{{\rm exp}}
\def\max{{\rm max}}

\def\Tr{{\rm Tr}}

\def\Bell{{\rm Bell}}
\def\ch{{\rm ch}}
\def\in{{\rm in}}
\def\out{{\rm out}}
\def\eff{{\rm eff}}
\bibliography{plain}
\title{Entanglement dynamics and decoherence of an atom coupled to a dissipative cavity field}
\vspace{20mm}
\author{
  S. J. Akhtarshenas
\thanks{E-mail:akhtarshenas@phys.ui.ac.ir}
, M. Khezrian
\thanks{E-mail:khezrian.m@gmail.com}
\\
{\small Department of Physics, University of Isfahan, Isfahan,
Iran } \\
{\small Quantum Optics Group, University of Isfahan, Isfahan, Iran
}}
\begin{document}
\maketitle

\begin{abstract}
In this paper, we investigate the entanglement dynamics and
decoherence in the interacting system of a strongly driven
two-level atom and a single mode vacuum field in the presence of
dissipation for the cavity field. Starting with an initial product
state with the atom in a general pure state and the field in a
vacuum state, we show that the final density matrix is supported
on ${\mathbb C}^2\otimes{\mathbb C}^2$ space,  and therefore, the
concurrence can be used as a measure of entanglement between the
atom and the field.  The influences of the cavity decay on the
quantum entanglement of the system are also discussed. We also
examine the Bell-CHSH violation between the atom and the field and
show that there are entangled states for which the Bell-BCSH
inequality is not violated. Using the above system as a quantum
channel, we also investigate the quantum teleportation of a
generic qubit state and also a two-qubit entangled state, and show
that in both cases the atom-field entangled state can be useful to
teleport an unknown state with fidelity better than any classical
channel.


\end{abstract}
\section{Introduction}
Quantum entanglement is one of the most prominent nonclassical
properties of quantum mechanics which has recently attracted much
attention in view of its connection with the theory of quantum
information and computation. The rapidly increasing in quantum
information processing has stimulated the interest of studying the
quantum entanglement. It has been recognized that entanglement
provides a fundamental potential resource for communication and
information processing \cite{Bennett1,Bennett2,Bennett3} and it
is, therefore, essential to create and manipulate entangled states
for quantum information application. Entanglement is usually
arising from quantum correlations between separated subsystems
which can not be created by local actions on each subsystems. A
pure quantum state of two or more subsystems is said to be
entangled if it is not a product of states of each components. On
the other hands, a bipartite mixed state $\rho$ is said to be
entangled if it can not be expressed as a convex combination of
pure product states \cite{Werner}, otherwise, the state is
separable or classically correlated.

Entangled states are very fragile when they are exposed to
environment. Actually, the biggest enemy of entanglement is
decoherence which is believed to be the responsible mechanism for
emergence of the classical behavior in quantum systems
\cite{Kofler,Joos}. Since the maintenance and control of entangled
states is essential to realization of quantum information
processing systems, the study of deteriorating effect of
decoherence in entangled states would be of considerable
importance from theoretical  as well as experimental  point of
view \cite{Dur,Carvalho,Simon}.

Entanglement dynamics and decoherence have been studied in the
frame of various models. The interaction of a two-level atom with
a single mode of the electromagnetic field, described by the
Jaynes-Cummings model \cite{Jaynes}, is one of the simplest and
most fundamental quantum systems. The Jaynes-Cummings model and
related models with dissipation have more recently attracted
interest in studies of quantum entanglement
\cite{Peixoto1999,Zhou2001,Sanz2002,Rendel2003,Dodonov2003,Li2004,Tanas2004,Can2004,Banacloche2005,Janowicz2006,Naderi2006,Li2008}.
Solano et al \cite{Solano2003} have shown that multipartite
entanglement can be generated by putting several two-level atoms
in a cavity of high quality factor with a strong classical driving
field. The resonance interaction of a cavity mode with a two-level
atom that is driven by a coherent field have considered by
Casagrande et al \cite{Casagrande2006}. They have shown that the
system can reach the maximum entanglement after a unitary
evolution for long enough interaction times.  Lougovski et al
\cite{Lougo2007} have proposed the implementation of a strongly
driven one-atom laser, based on the off-resonant interaction of a
three-level atom in $\Lambda$ configuration with a single cavity
mode and three laser fields. They have shown that the system can
be well approximated by a two-level atom resonantly coupled to the
cavity mode and driven by a strongly effective coherent field.
They have also studied the entanglement properties of the
atom-field system on a time scale much shorter than the cavity
decay time, where the atom-field system is almost a pure state.
The entanglement of an open tripartite system where a cavity field
mode in thermal equilibrium is off-resonantly coupled with two
atoms that are simultaneously driven by a resonant coherent field
have investigated in \cite{Casagrande2008EPJD}. Bina et al
\cite{Bina2008} have studied entanglement between two strongly
driven atoms resonantly coupled to a dissipative cavity field
mode. They have shown that for this system,  the master equation
is analytically solvable. In Ref. \cite{Bina2008EPJD}, the authors
have studied  the dynamics of an open quantum system where $N$
strongly driven two-level atoms are equally coupled on resonance
to a dissipative cavity mode, and have shown that also in this
case the master equation is analytically solvable. Very recently,
Zhang et al \cite{Zhang2009} have investigated entanglement
dynamics and purity of a two-level atom, driven by a classical
field, and interacting with a coherent field in a dissipative
environment.

The aim of our paper is to analyze the dynamics of the
Jaynes-Cummings model in order to find relation between the
entanglement of the atomic and the filed degrees of freedom and
the decoherence in the presence of dissipation for the field. The
system considered here consists of a strongly driven two-level
atom resonantly coupled to a dissipative cavity field mode
\cite{Lougo2007}. We start initially with the atom in a general
pure state and the field in a vacuum state and show that the final
density matrix is supported on ${\mathbb C}^2\otimes{\mathbb C}^2$
space, and therefore the concurrence can be used as a measure of
the degree of entanglement between the atom and the field. The
influences of the cavity decay on the quantum entanglement of the
system are investigated, and find that the dissipation  suppresses
the entanglement. We also examine the Bell-CHSH violation between
the atom and the field and show that there are entangled states
for which the Bell-BCSH inequality is not violated.  The
decoherence induced by the cavity is also studied and it is shown
that the coherence properties of the atom and also the field are
affected by the cavity. The possibility of writing the atom-field
density matrix as a two-qubit system enables us to use the
atom-field system as a quantum channel for teleportation.  The
one-qubit teleportation and also the two-qubit entanglement
teleportation via the quantum channel constructed by the
atom-field system are also investigated and the fidelity of the
teleportation and also the entanglement of the replica are also
discussed. We show that in both cases the atom-field entangled
state can be useful to teleport an unknown state with fidelity
better than any classical channel.

The paper is organized as follows: In section 2, we introduce the
Hamiltonian of an atom interacting with  a single mode vacuum
field in the presence of dissipation. We also give the solution of
the master equation in this section. In section 3, we study
entanglement of the atom-field system by using the concurrence,
and investigate the effect of dissipation on the concurrence. We
also examine the possible  violation of the Bell-CHSH inequality.
Section 4 is devoted to investigating  the effect of dissipation
on the purity of the system and its corresponding subsystems. The
possibility of using the entanglement between the atom and the
field as a resource to teleport the one-qubit and two-qubit states
is also considered in section 5.   The paper is concluded in
section 6 with a brief conclusion.

\section{Master equation and solution }
The starting point for our analysis is the following Hamiltonian
for the atom-field interaction \cite{Bina2008}
\begin{equation}
 \hat{H}(t)=\frac{\hbar\omega_{a}}{2}\hat{\sigma}_{z}+\hbar\omega_{f}\hat{a}^{\dag}\hat{a}+\hbar\Omega
\left(e^{-i\omega_{D}t}\hat{\sigma}^{\dag}+e^{i\omega_{D}t}\hat{\sigma}\right)
+\hbar
g\left(\hat{\sigma}^{\dag}\hat{a}+\hat{\sigma}\hat{a}^{\dag}\right).
\end{equation}
This Hamiltonian describes a driven two-level atom interacting
with a cavity field. Here $g$ is the atom-field coupling constant,
$\Omega$ is the Rabi frequency associated with the coherent
driving field amplitude,
$\omega_{a}=(\epsilon_{e}-\epsilon_{g})/\hbar$ is the atomic
transition frequency, $\omega_{f}$ denotes the field frequency,
and $\omega_D$ is the frequency of the classical field. The atomic
"spin-flip" operators $\hat{\sigma}=|g\rangle\langle e|
(\hat{\sigma}^{\dag}=|e\rangle\langle g|)$, and the atomic
inversion operator $\hat{\sigma}_{z}=|e\rangle\langle
e|-|g\rangle\langle g|$ act on the atom Hilbert space ${\mathcal
H}^{A}={\mathbb C}^{2}$ spanned by the excited state
$|e\rangle\rightarrow(1,0)^{T}$ and the ground state
$|g\rangle\rightarrow(0,1)^{T}$. The field annihilation and
creation operators $\hat{a}$ and $\hat{a}^{\dag}$ satisfy the
commutation relation $[\hat{a},\hat{a}^{\dag}]=1$ and act on the
field Hilbert space ${\mathcal H}^{F}$ spanned by the
photon-number states $\left\{|n\rangle
=\frac{(\hat{a}^{\dag})^{n}}{\sqrt{n!}}|0\rangle\right\}_{n=0}^{\infty}$.

In the following, we consider the dissipative dynamics for the
cavity field when it is in contact with the environment, but we
neglect  atomic decays. The dynamics of the atom-field density
operator $\hat{\rho}'$ is described by the master equation
\begin{equation}\label{MEq}
\dot{\hat{\rho}}'=-\frac{i}{\hbar}[\hat{H},\hat{\rho}']
+\hat{\mathcal{L}}_{f}\hat{\rho}',
\end{equation}
where the super-operator $\hat{\mathcal{L}}_{f}$ describes the
losses inside the cavity, and at zero temperature it is written as
follows
\begin{equation}
 \hat{\mathcal{L}}_{f}\hat{\rho}' =\frac{k}{2}\left[2\hat{a}\hat{\rho}'\hat{a}^{\dag}
-\hat{a}^{\dag}\hat{a}\hat{\rho}'-\hat{\rho}'\hat{a}^{\dag}\hat{a}\right],
\end{equation}
where $k$ is the cavity decay rate. In the interaction picture the
master equation (\ref{MEq}) can be written as
\begin{equation}
\dot{\hat{\rho}}_{I}=-\frac{i}{\hbar}[\hat{H}_{I},\hat{\rho}_{I}]
+\hat{\mathcal{L}}_{f}\hat{\rho}_{I}.
\end{equation}
The dissipative term remains unchanged, and the time-independent
Hamiltonian is given by $\hat{H}_{I}=\hat{H}_{0}+\hat{H}_{1}$ with
\begin{equation}
 \hat{H}_{0} =
 -\hbar\delta\hat{a}^{\dag}\hat{a}+\hbar\Omega(\hat{\sigma}^{\dag}+\hat{\sigma}),
 \qquad \hat{H}_{1} = \hbar
 g(\hat{\sigma}^{\dag}\hat{a}+\hat{\sigma}\hat{a}^{\dag}),
\end{equation}
where we introduced the atom-cavity field detuning parameter
$\delta=\omega_{a}-\omega_{f}$. Employing the unitary
transformation  $ \hat{U}(t)=\exp \left
\{\frac{i}{\hbar}\hat{H}_{0}t\right\}$, we arrive at the following
master equation for the density operator
$\hat{\rho}(t)=\hat{U}(t)\hat{\rho}_{I}(t)\hat{U}^{\dag}(t)$
\begin{equation}
\dot{\hat{\rho}}=-\frac{i}{\hbar}[\hat{U}\hat{H}_{1}\hat{U}^{\dag},\hat{\rho}]
+\hat{\mathcal{L}}_{f}\hat{\rho},
\end{equation}
with
\begin{equation}
\hat{U}\hat{H}_{1}\hat{U}^{\dag}=\frac{\hbar
g}{2}\left[|+\rangle\langle+|-|-\rangle\langle-|+e^{2i\Omega
t}|+\rangle\langle-|-e^{-2i\Omega
t}|-\rangle\langle+|\right]\hat{a} e^{i\delta t}+{\rm H.C},
\end{equation}
where H.C. stands for Hermitian conjugate and the rotated basis
$\{|+\rangle, |-\rangle\}$ is defined by
$|\pm\rangle=\frac{1}{\sqrt{2}}(|g\rangle\pm|e\rangle)$. On
resonance $(\delta=0)$, and in the strong-driving regime for
interaction between the atom and the external field, $\Omega\gg
g$, and in the rotating-wave approximation, the following
effective master equation is obtained
\begin{equation}\label{MasterEq}
\dot{\hat{\rho}}=-\frac{i}{\hbar}[\hat{H}_{\eff}(t),\hat{\rho}]
+\hat{\mathcal{L}}_{f}\hat{\rho},
\end{equation}
with the effective Hamiltonian
\begin{equation}\label{H}
 \hat{H}_{\eff}=\frac{\hbar
 g}{2}(\hat{\sigma}^{\dag}+\hat{\sigma})(\hat{a}+\hat{a}^{\dag}).
\end{equation}
Hamiltonian (\ref{H}) contains both the Jaynes-Cummings term
$(\hat{\sigma}^{\dag}\hat{a}+\hat{\sigma}\hat{a}^{\dag})$ and the
anti-Jaynes-Cummings term
$(\hat{\sigma}^{\dag}\hat{a}^{\dag}+\hat{\sigma}\hat{a})$
\cite{Lougo2007}. In the following we will describe the solution
of the above effective master equation.


In order to solve the master equation (\ref{MasterEq}), we follow
the method introduced in \cite{Lougo2007,Bina2008}. Let us first
introduce the following decomposition for the density operator
$\hat{\rho}(t)$ of the whole system
\begin{equation}
 \hat{\rho}(t)=\sum_{i,j=1}^{2}\langle
 i|\hat{\rho}(t)|j\rangle|i\rangle\langle j|
 =\sum_{i,j=1}^{2}\hat{\rho}_{ij}|i\rangle\langle j|,
\end{equation}
where $\{|i\rangle\}_{i=1,2}=\{|+\rangle,|-\rangle\}$ is the
rotated basis of the atom, and $\hat{\rho}_{ij}(t)=\langle
i|\hat{\rho}(t)|j\rangle$ are operators acting on the field
Hilbert space. With this definition, the master equation
(\ref{MasterEq}) is equivalent to the following set of uncoupled
equations for the field operators $\hat{\rho}_{ij}(t)$
\begin{eqnarray}\label{2}
 \nonumber
  \dot{\hat{\rho}}_{11} &=&
  -\frac{i g}{2}\left[\hat{a}^{\dag}+\hat{a},\hat{\rho}_{11}\right]
  +\frac{k}{2}\left(2\hat{a}\hat{\rho}_{11}\hat{a}^{\dag}
-\hat{a}^{\dag}\hat{a}\hat{\rho}_{11}-\hat{\rho}_{11}\hat{a}^{\dag}\hat{a}\right),
  \\
  \dot{\hat{\rho}}_{12} &=& -\frac{i g}{2}\left\{\hat{a}^{\dag}+\hat{a},\hat{\rho}_{12}\right\}
  +\frac{k}{2}\left(2\hat{a}\hat{\rho}_{12}\hat{a}^{\dag}
-\hat{a}^{\dag}\hat{a}\hat{\rho}_{12}-\hat{\rho}_{12}\hat{a}^{\dag}\hat{a}\right),\\
  \nonumber
  \dot{\hat{\rho}}_{22} &=& \frac{i g}{2}\left[\hat{a}^{\dag}+\hat{a},\hat{\rho}_{22}\right]
  +\frac{k}{2}\left(2\hat{a}\hat{\rho}_{22}\hat{a}^{\dag}
-\hat{a}^{\dag}\hat{a}\hat{\rho}_{22}-\hat{\rho}_{22}\hat{a}^{\dag}\hat{a}\right),
\end{eqnarray}
where  $\{ , \}$ denotes  anti-commutator symbol and
$\dot{\hat{\rho}}_{21}(t)=[\dot{\hat{\rho}}_{12}(t)]^{\dag}$. Now
following the method of reference \cite{Bina2008}, let us first
define the following functions in the phase space associated with
the field
\begin{equation}\label{chi}
  \chi_{ij}(\beta,t)=Tr_{f}\left[\hat{\rho}_{ij}(t)\hat{D}(\beta)\right],
  \qquad
  \hat{D}(\beta)=\exp\left[\beta\hat{a}^{\dag}-\beta^{*}\hat{a}\right].
\end{equation}
In this representation, the equations (\ref{2}) take the following
from
\begin{eqnarray}\label{chitEq}
\nonumber
 \dot{\chi}_{11}(\beta,t) &=& \frac{i g}{2}(\beta+\beta^{*})\chi_{11}(\beta,t)
 -\frac{k}{2}\left(\beta\frac{\partial}{\partial\beta}+
 \beta^{*}\frac{\partial}{\partial\beta^{*}}+\left|\beta\right|^{2}\right)\chi_{11}(\beta,t),\\
 \dot{\chi}_{12}(\beta,t) &=&
 -ig\left[\frac{\partial}{\partial\beta}
 -\frac{\partial}{\partial\beta^{*}}\right]\chi_{12}(\beta,t)
 -\frac{k}{2}\left(\beta\frac{\partial}{\partial\beta}+
 \beta^{*}\frac{\partial}{\partial\beta^{*}}+\left|\beta\right|^{2}\right)\chi_{12}(\beta,t),\\
 \nonumber
 \dot{\chi}_{22}(\beta,t) &=& -\frac{i g}{2}(\beta+\beta^{*})\chi_{22}(\beta,t)
 -\frac{k}{2}\left(\beta\frac{\partial}{\partial\beta}+
 \beta^{*}\frac{\partial}{\partial\beta^{*}}+\left|\beta\right|^{2}\right)\chi_{22}(\beta,t).
\end{eqnarray}
We now assume that at $t=0$ the atom is described by the pure
state $|\psi_{a}(0)\rangle =\cos{\theta/2}|
+\rangle+\e^{i\phi}\sin{\theta/2}|-\rangle$, with
$0\leq\theta\leq\pi,\;\;0\leq\phi<2\pi$, and the cavity field is
in the the vacuum state $|\psi_{f}(0)\rangle=|0\rangle$. In the
representation given by (\ref{chi}), the above initial state takes
the following form
\begin{eqnarray}\nonumber
 \chi_{11}(\beta,0)&=&\cos^2{\theta/2}\;\exp\left(-|\beta|^{2}/2\right),
\\
 \chi_{12}(\beta,0)&=&\frac{1}{2}\e^{-i\phi}\sin{\theta}\;\exp\left(-|\beta|^{2}/2\right),
\\  \nonumber
 \chi_{22}(\beta,0)&=&\sin^2{\theta/2}\;\exp\left(-|\beta|^{2}/2\right),
\end{eqnarray}
and $\chi_{21}(\beta,0)=\chi_{12}^{\ast}(\beta,0)$. Now under the
above initial conditions, the equations (\ref{chitEq}) can be
solved by using the method of characteristics \cite{BarnettBOOK},
and we get
\begin{eqnarray}
\nonumber
 \chi_{11}(\beta,t) &=& \cos^{2}{\theta/2}\;\exp\left(-\frac{|\beta|^{2}}{2}-
 \alpha^{*}(t)\beta+\alpha(t)\beta^{*}\right),\\
 \chi_{12}(\beta,t) &=& \frac{1}{2}\e^{-i\phi}\sin{\theta}
 \;f(t)\;\exp\left(-\frac{|\beta|^{2}}{2}+
 \alpha^{*}(t)\beta+ \alpha(t)\beta^{*}\right),\\ \nonumber
 \chi_{22}(\beta,t) &=& \sin^{2}{\theta/2}\;\exp\left(-\frac{|\beta|^{2}}{2}+
 \alpha^{*}(t)\beta-\alpha(t)\beta^{*}\right),
\end{eqnarray}
and $\chi_{21}(\beta,t)=\chi_{12}^{\ast}(\beta,t)$. In the above
equations we have defined the time dependent coherent field
amplitude $\alpha(t)$ and the function $f(t)$ as
\begin{equation}\label{alphaf}
\alpha(t)=i\frac{g}{k}\left(1-e^{-kt/2}\right), \quad
f(t)=\exp\left(-2\left(\frac{g}{k}\right)^2kt
+4\left(\frac{g}{k}\right)^2\left(1-e^{-kt/2}\right)\right).
\end{equation}
From the above expressions we find
\begin{eqnarray}\label{element dencity}
\nonumber
 \hat{\rho}_{11}(t) &=& \cos^{2}{\theta/2}
 \left|-\alpha(t)\right\rangle\left\langle-\alpha(t)\right|,\\
\nonumber
 \hat{\rho}_{12}(t) &=& \frac{1}{2}\;\e^{-i\phi}\sin{\theta}
 \;f(t)\e^{2\left|\alpha(t)\right|^{2}}
 \left|-\alpha(t)\right\rangle\left\langle\alpha(t)\right|,\\
 \hat{\rho}_{22}(t) &=& \sin^{2}{\theta/2}
 \left|\alpha(t)\right\rangle\left\langle\alpha(t)\right|,
\end{eqnarray}
and $\hat{\rho}_{21}(t)=\hat{\rho}_{12}^{\dag}(t)$. As a matter of
fact, by choosing the initial state of the atom as
$\theta=\pi/2,\; \phi=0$, the above density matrix reduces to the
relation (31) of Ref. \cite{Lougo2007}. However our objective here
is to study the effect of dissipation on the entanglement of the
atom-field system and also the possibility  of using  this system
as a quantum channel for efficient quantum teleportation.

\section{Quantum entanglement}
In what follows, we will study entanglement dynamics of the above
state.  To some extent, the dynamics of entanglement is the time
evolution of entanglement measures. Many entanglement measures
have been introduced and analyzed in the literature, but the one
most relevant to this work is entanglement of formation, which in
fact intends to quantify the resources needed to create a given
entangled state \cite{Bennett3}. Remarkably, Wootters
\cite{Wootters} has shown that entanglement of formation of a
two-qubit state $\hat{\rho}$ is related to a quantity called
concurrence as
\begin{equation}
E(\hat{\rho})=\Xi[C(\hat{\rho})]=h\left(\frac{1}{2}+\frac{1}{2}\sqrt{1-C^2}\right),
\end{equation}
where $h(x)=-x \log_2 x-(1-x)\log_2(1-x)$ is the binary entropy
function and $C(\hat{\rho})$ is the concurrence of the state
$\hat{\rho}$, defined by
\begin{equation}\label{concurrence}
 C(\hat{\rho})=\max\left\{0,
 \sqrt{\lambda_{1}}-\sqrt{\lambda_{2}}
 -\sqrt{\lambda_{3}}-\sqrt{\lambda_{4}}\right\},
\end{equation}
where the $\lambda_{i}$ are the non-negative eigenvalues, in
decreasing order, of the non-Hermitian matrix $\hat{\rho}{\tilde
{\hat{\rho}}}$. Here $\tilde{\hat{\rho}}$ is the matrix given by
$\tilde{\hat{\rho}}=\left(\sigma_{y}\otimes\sigma_{y}\right)\hat{\rho}^{*}
 \left(\sigma_{y}\otimes\sigma_{y}\right)$
where $\hat{\rho}^{*}$ is the complex conjugate of $\hat{\rho}$
when it is expressed in a standard basis such as
$\{\left|11\right>, \left|12\right>,\left|21\right>,
\left|22\right>\}$ and $\sigma_y$ represents Pauli matrix in the
local basis $\{\left|1\right>, \left|2\right>\}$. Furthermore, the
function $\Xi$ is a monotonically increasing function of the
concurrence $C(\hat{\rho})$, and ranges from 0 to 1 as
$C(\hat{\rho})$ goes from 0 to 1, so that one can take the
concurrence as a measure of entanglement in its own right.

Equation (\ref{element dencity}) shows that the operators
$\hat{\rho}_{ij}(t)$ act on a field subspace spanned by two
vectors $|\alpha(t)\rangle $ and $|-\alpha(t)\rangle$. Now it is
easy to see that the determinant obtained from the inner product
of these two vectors is equal to $1-\e^{-4|\alpha(t)|^2}$ which is
nonzero provided that $|\alpha(t)|\ne 0$. This means that two
vectors $|\alpha(t)\rangle $ and $|-\alpha(t)\rangle$ are linearly
independent provided that $t\ne 0$ and $g\ne 0$. Therefore the
final density matrix $\hat{\rho}(t)$ is supported at most on
${\mathbb C}^2\otimes{\mathbb C}^2$ space, and thus, one can use
the concurrence as a measure of entanglement between the atom and
the field.  Now in order to calculate the concurrence for the
atom-field density matrix given in Eq. (\ref{element dencity}), we
must first write the density matrix in an  orthonormal product
basis. To this aim, we use the Gram-Schmidt procedure
\cite{Nielsen2000} to construct two orthonormal vectors
$|v_{1}\rangle$ and $|v_{2}\rangle$ as
\begin{equation}
  |v_{1}\rangle = \left|\alpha(t)\right\rangle, \qquad
  |v_{2}\rangle = \frac{\left|-\alpha(t)\right\rangle
  -x(t)\left|\alpha(t)\right\rangle}
  {\sqrt{1-x^2(t)}},
\end{equation}
where
\begin{equation}
 x(t) =\langle \alpha(t)|-\alpha(t)\rangle= \exp\left(-2\left|\alpha(t)\right|^{2}\right).
 \end{equation}
Two vectors $|v_{1}\rangle$ and $|v_{2}\rangle$ span, effectively,
the space of the field and constitute the field qubit states.
Therefore, in our model, the atom-field system constitute a
two-qubit system. Now in the orthonormal basis
$\{|+\rangle|v_1\rangle,|+\rangle|v_2\rangle,|-\rangle|v_1\rangle,|-\rangle|v_2\rangle\}$,
the atom-field density matrix can be represented by
\begin{equation}\label{RhoT}
   \hat{\rho}(t)=
   \begin{pmatrix}
    x^2(t)\cos^{2}{\theta/2} & x(t)\sqrt{1-x^{2}(t)}\cos^{2}{\theta/2}
    & \frac{1}{2}f(t)\sin{\theta}\;
    \e^{-i\phi} & 0 \\
    x(t)\sqrt{1-x^{2}(t)}\cos^{2}{\theta/2} & (1-x^{2}(t))\cos^{2}{\theta/2}
    & \frac{\sqrt{1-x^{2}(t)}}{2x(t)}f(t)\sin{\theta}
    \;\e^{-i\phi} & 0 \\
    \frac{1}{2}f(t)\sin{\theta}
    \;\e^{i\phi} & \frac{\sqrt{1-x^{2}(t)}}{2x(t)}f(t)\sin{\theta}
    \;\e^{i\phi}
    &\sin^{2}{\theta/2} & 0 \\
     0 & 0 & 0 & 0 \
   \end{pmatrix}.
\end{equation}
Now we can use the concurrence as a measure of entanglement
between the atom and the field. For the atom-field state defined
in equation (\ref{RhoT}) we obtain
\begin{eqnarray}
\nonumber
  \lambda_{1}&=&\frac{(1-x^{2}(t))}{4x^{2}(t)}(x(t)+f(t))^{2}\sin^{2}{\theta},\\
\nonumber
  \lambda_{2}&=&\frac{(1-x^{2}(t))}{4x^{2}(t)}(x(t)-f(t))^{2}\sin^{2}{\theta},\\
  \lambda_{3}&=&\lambda_{4}=0,
\end{eqnarray}
and therefore the concurrence between the atom and the field is
given by
\begin{equation}\label{C(t)}
 C(t)=\max\left\{0,\frac{\sqrt{1-x^{2}(t)}}{x(t)}f(t)\sin{\theta}
 \right\}.
\end{equation}
It is clear that for $\theta=0$  the concurrence is zero for all
times, i.e. the atom described by the initial state
$|+\rangle=\frac{1}{\sqrt{2}}(|g\rangle+|e\rangle)$ does not get
entangled with the field.  Indeed, in this case the final state of
the system is described by the pure state
$\rho(t)=|\psi(t)\rangle\langle \psi(t)|$, where
$|\psi(t)\rangle=|+\rangle|-\alpha(t)\rangle$. This means that the
initial state $|+\rangle|0\rangle$ of the system  defines a
decoherence-free subspace in which the time evolution of the system
is unitary. Thus the state $|+\rangle|0\rangle$ does not become
entangled with the environment. But  in the absence of dissipation,
i.e. k=0, the unitary evolution operator
$\hat{U}=\exp\{-i\hat{H}_{\eff}t/\hbar\}$ of the Hamiltonian
(\ref{H}) can be written as \cite{Lougo2004}
\begin{equation}
\hat{U}(\xi(t))=|+\rangle\langle
+|\hat{D}(-\xi(t))+|-\rangle\langle -|\hat{D}(\xi(t)), \qquad
\xi(t)=igt/2
\end{equation}
where $\hat{D}$ is the displacement operator defined in Eq.
(\ref{chi}). The evolution of the initial state
$|+\rangle|0\rangle$  by the above unitary operator leads to the
product state  $|+\rangle|-\xi(t)\rangle$. Motivated by this we
can say that in the decoherence-free subspace the unitary
evolution of the system is governed by the unitary operator
$\hat{U}(\alpha(t))$ where $\alpha(t)$ is the coherent field
amplitude defined in Eq. (\ref{alphaf}). Obviously, for $k=0$ we
have $\alpha(t)=\xi(t)$.
\begin{figure}[t]
\centerline{\includegraphics[height=9cm]{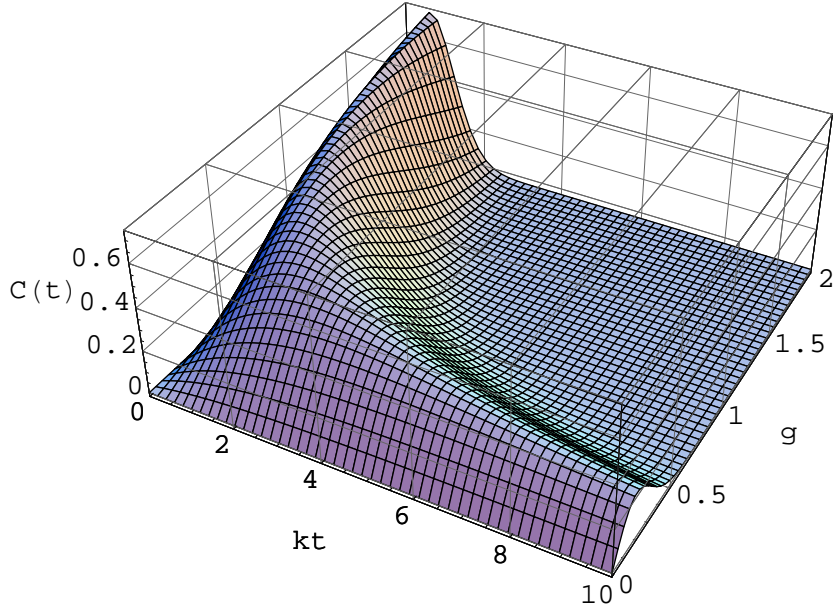}}
\caption{Concurrence $C(t)$ is plotted as a function of $kt$ and
coupling constant $g$ with $\theta=\pi/2$.}
\end{figure}
\begin{figure}[t]
\centerline{\includegraphics[height=11cm]{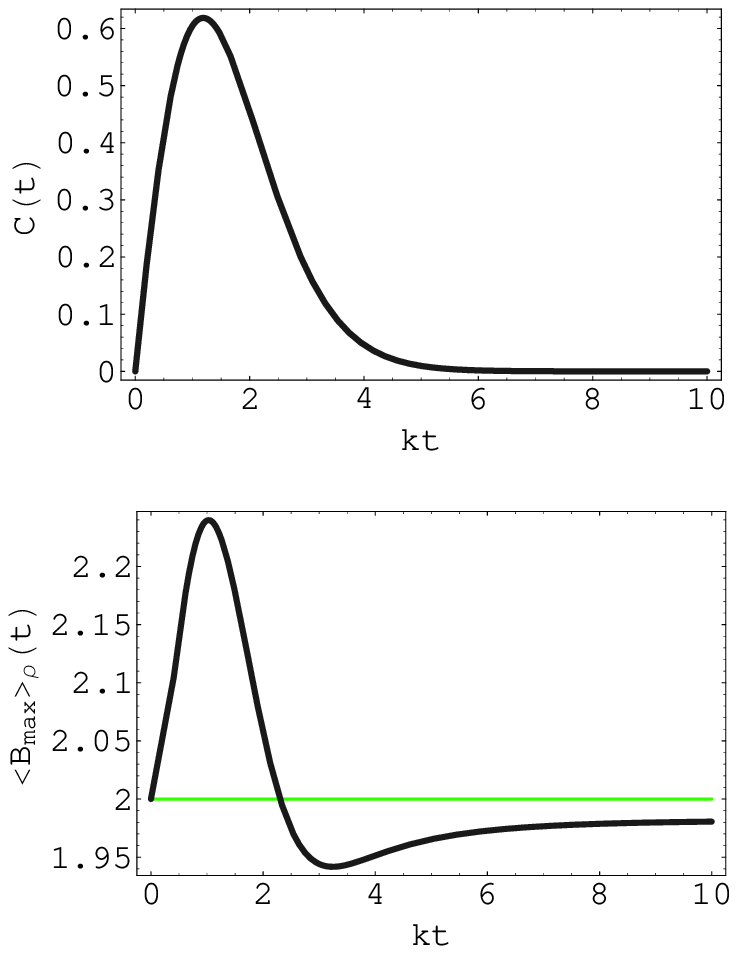}}
\caption{Concurrence $C(t)$ (upper panel) and the maximal violation
measure $\langle{\mathcal B}_{\max}\rangle_{\hat{\rho}}(t)$ (lower
panel) are plotted as a function of $kt$ with $\theta=\pi/2$ and
$g=1$. The horizontal line in the lower figure shows the minimum
violation.}
\end{figure}
In order to show the effect of dissipation rate $k$ and coupling
constant $g$ on the entanglement of the system, we plot the
concurrence as a function of $kt$ and the coupling constant $g$ in
Fig. 1. It shows that the entanglement of the system increases with
increase of the coupling constant $g$, and decreases with increase
of the dissipation rate $k$. Furthermore, the asymptotic long time
density matrix is separable and has the following form.
\begin{equation}\label{RhoLongTime}
 \hat{\rho}(\infty)=\cos^{2}{\theta/2}|+\rangle\langle+|\otimes
 |-\alpha(t)\rangle\langle-\alpha(t)|+\sin^{2}{\theta/2}|-\rangle\langle-|\otimes
 |\alpha(t)\rangle\langle\alpha(t)|.
\end{equation}

Now in the following, we attempt to discuss nonlocality of the atom
and the field.  The most commonly discussed Bell inequality is the
Clauser-Horne-Shimony-Holt (CHSH) inequality \cite{CHSH}. The
Bell-CHSH operator formulated for two-qubit systems  has the
following form \cite{CHSH}
\begin{equation}\label{CHSHOperator}
{\mathcal B}=\bf{a}\cdot
\bf{\sigma}\otimes(\bf{b}+\bf{b}^{\prime})\cdot\bf{\sigma}+\bf{a}^{\prime}\cdot
\bf{\sigma}\otimes(\bf{b}-\bf{b}^{\prime})\cdot\bf{\sigma},
\end{equation}
where $\bf{a}, \bf{a}^{\prime},\bf{b}, \bf{b}^{\prime}$ are unite
vectors in ${\mathbb R}^3$ and $\{\sigma_i\}_{i=1}^{3}$ are the
standard Pauli matrices. The Bell-CHSH inequality  states that
within any local model the expectation value $\langle {\mathcal
B}\rangle_{\hat{\rho}}\equiv\Tr{(\hat{\rho}{\mathcal B})}$ of the
Bell-CHSH operator has to be bounded by 2, i.e.
\begin{equation}\label{CHSHInequality}
|\langle {\mathcal B}\rangle_{\hat{\rho}}|\le 2.
\end{equation}
Horodecki et al have presented an effective criterion for
violating the Bell-CHSH inequality by an arbitrary mixed two-qubit
state \cite{Horodecki95}. They have shown that the maximum amount
of Bell violation of a two-qubit state $\hat{\rho}$, i.e.
$\langle{\mathcal B}_{\max}\rangle_{\hat{\rho}}=\max_{\mathcal
B}^{}|\langle {\mathcal B}\rangle_{\hat{\rho}}|$,  is given by
$2\sqrt{\mu+{\tilde \mu}}$ where $\mu,{\tilde \mu}$ are two
greater eigenvalues of the matrix
$T_{\hat{\rho}}^{\dag}T_{\hat{\rho}}$. Here the matrix
$T_{\hat{\rho}}$ is a $3\times 3$ matrix whose elements are
$[T_{\hat{\rho}}]_{ij}=\Tr{(\hat{\rho}\sigma_i\otimes \sigma_j)}$,
and is responsible for correlations. It follows, therefore, from
this maximal violation measure that a state shows Bell violation
when $\langle{\mathcal B}_{\max}\rangle_{\hat{\rho}}>2$ and the
maximal violation when $\langle{\mathcal
B}_{\max}\rangle_{\hat{\rho}}=2\sqrt{2}$.

Now, it is not difficult to see that for the atom-field density
operator $\hat{\rho}(t)$ given by equation (\ref{RhoT}), the
maximal violation measure can be written as
\begin{equation}
\langle{\mathcal
B}_{\max}\rangle_{\hat{\rho}}(t)=2\sqrt{1+\left(\frac{f^2(t)}{x^2(t)}-x^2(t)\right)\sin^2{\theta}}.
\end{equation}
In Fig. 2 we plot the concurrence $C(t)$ (upper panel) and the
maximal violation measure $\langle{\mathcal
B}_{\max}\rangle_{\hat{\rho}}(t)$ (lower panel) as a function of
$kt$ with $\theta=\pi/2$ and $g=1$. The horizontal line in the
lower figure shows the boundary value 2 in equation
(\ref{CHSHInequality}), i.e. the  minimum violation. There we can
clearly see that there are entangled states for which the
Bell-BCSH inequality is not violated. It is worth  noting that
although the atom-field entanglement disappears asymptotically,
but the nonlocality defined  by the Bell-CHSH inequality
disappears at a finite time.

\section{Decoherence}
In quantum information processing, decoherence is another
essential problem that deserves some attention.  Generally,
decoherence is used to estimate the deviation from an ideal state
and can be considered as a symbol to express the reduction of
purity and, therefore, one can use the linear entropy
$S(\;\hat{\rho})=1-\Tr\;[\;\hat{\rho}^{2}\;]$ as a measure of
decoherence. The linear  entropy has the limiting values 0 and
$1-1/N$, respectively, for pure and maximally mixed states, where
$N$ is the dimension of the space that the density matrix
$\hat{\rho}$ is supported on. The linear entropy of the atom-field
system is given by
\begin{equation}
 S(\;\hat{\rho})=1-\Tr\;[\;\hat{\rho}^{2}\;]=\frac{1}{2}\;\left(1-\frac{f^{2}(t)}{x^2(t)}\right)\sin^{2}\theta.
\end{equation}
On the other hand, the reduced density matrix of the atom can be
obtained by tracing out over the field degrees of freedom where we
get
\begin{eqnarray}
  \hat{\rho}_{a}(t)=\Tr_{f}[\;\hat{\rho}(t)\;]=
 \begin{pmatrix}
    \cos^{2}{\theta/2}&
    \frac{1}{2}\;\e^{-i\phi}\sin{\theta}\;f(t)\\
    \frac{1}{2}\;\e^{i\phi}\sin{\theta}\;f(t) &
    \sin^{2}{\theta/2}\
  \end{pmatrix}.
\end{eqnarray}
The linear entropy of the atom is given by
\begin{equation}
 S(\;\hat{\rho}_{a})=1-\Tr\;[\;\hat{\rho}_{a}^{2}\;]=\frac{1}{2}\;(1-f^{2}(t))\sin^{2}\theta.
\end{equation}
Similarly, we can obtain the reduced density matrix of the field
by tracing out over the atom degrees of freedom and get
\begin{eqnarray}
  \hat{\rho}_{f}(t)=\Tr_{a}[\;\hat{\rho}(t)\;]=\begin{pmatrix}
    x^2(t)\cos^{2}{\theta/2}+\sin^{2}{\theta/2}&
    x(t)\sqrt{1-x^2(t)}\cos^{2}{\theta/2}\\
    x(t)\sqrt{1-x^2(t)}\cos^{2}{\theta/2}&
    (1-x^2(t))\cos^{2}{\theta/2}\
  \end{pmatrix},
\end{eqnarray}
where, clearly, shows that the field reduced density matrix does
not depend on the decoherence function $f(t)$.  This matrix can be
used to calculate the linear entropy of the field as
\begin{equation}
 S(\;\hat{\rho}_{f})=1-\Tr\;[\;\hat{\rho}_{f}^{2}\;]=
 \frac{1}{2}\;(1-x^2(t))\sin^{2}\theta.
\end{equation}

\begin{figure}[t]
\centerline{\includegraphics[height=8cm]{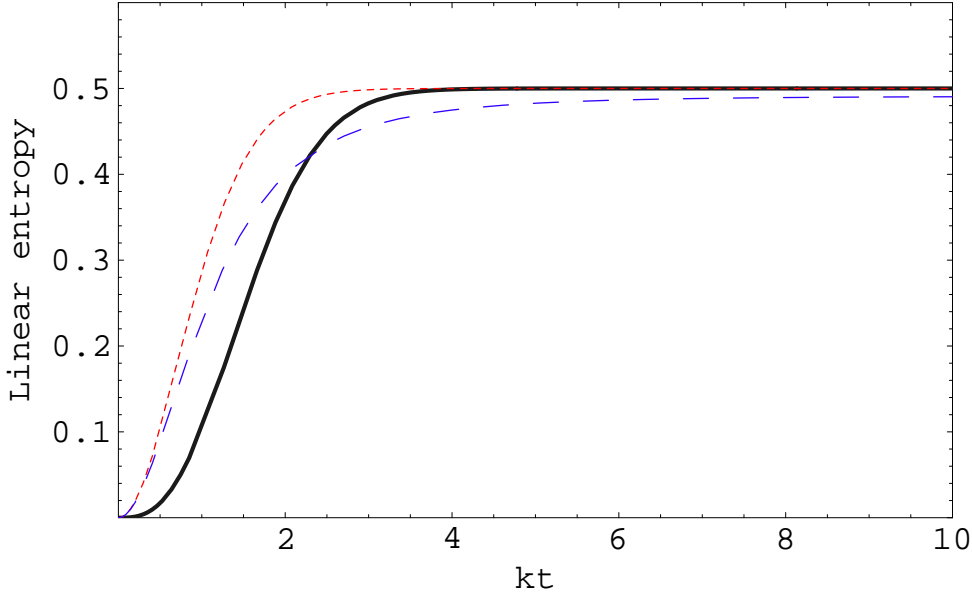}}
\caption{Linear entropy of atom-field (solid line), atom (dotted
line) and field (dashed line) are plotted as a function of $kt$ with
$\theta=\pi/2$ and $g=1$.}
\end{figure}
It is clear that all of the three linear entropies obtained above
are proportional to $\sin{\theta}$, and therefore when $\theta=0$
the atom-field system and its corresponding subsystems have zero
entropies. As we mentioned already this is because of the fact that
in this particular case the time evolution of the system is unitary
and the state remains separable as well as pure.  In Fig. 3 we plot
the linear entropy of the atom-field system and its two reduced
subsystems. It is clear that all three linear entropy have
asymptotic values near $1/2$. Although this asymptotic value for the
atom subsystem is the maximum value that the atom can gain, i.e.
$1/2$,  but for the field subsystem it is equal to $(1-\e^{-4})/2$.
Peixoto et al \cite{Peixoto1999} have employed the Jaynes-Cummings
model in the dispersive approximation for a dissipative cavity at
zero temperature and showed that the cavity has practically no
influence on the coherence properties of the field from the
qualitatively point of view, but the atom's coherence properties are
strongly influenced by dissipation both qualitatively and
quantitatively, although it is not directly coupled to the cavity.
But our results show that the coherence properties of the field are
also affected by the cavity.

\section{Teleportation }
An important aspect of quantum nonseparability is the quantum
teleportation, discovered by Bennett et al \cite{Bennett2}.
Bennett et al have shown that two spin-$\frac{1}{2}$ particles,
separated in space and entangled in a singlet state, can be used
for teleportation. Popescue \cite{Popescu} noticed that the pairs
in a mixed state could still be useful for (imperfect)
teleportation, but they reduce the fidelity of teleportation. It
has been shown  \cite{Popescu,Massar,Gisin} that the purely
classical channel can give at most fidelity $F=\frac{2}{3}$. The
possibility of using the entanglement between the atom and the
field as a resource for the standard teleportation protocol
${\mathcal P}_0$ is considered in the next subsection.

\subsection{One qubit teleportation}
The standard teleportation ${\mathcal P}_0$ \cite{Bennett2}
involves two particle sources producing pairs in a given mixed
state $\hat{\rho}_{\ch}$ which forms the quantum channel. This
quantum channel is equivalent to a generalized depolarizing
channel $\Lambda^{\hat{\rho}_{\ch},{\mathcal P}_0}$, with
probabilities given by the maximally entangled components of the
resources \cite{Bowen,Albeverio}. Now we look at the standard
protocol ${\mathcal P}_0$, using the atom-field state
$\hat{\rho}(t)$, i.e. a two-qubit mixed state, as resource. We
consider as an input state a one-qubit system in an unknown pure
state
$|\psi_{\in}\rangle=\cos{\vartheta/2}|+\rangle+e^{i\varphi}\sin{\vartheta/2}|-\rangle$
with $0\leq\vartheta\leq\pi,\;\;0\leq\varphi<2\pi$. The density
matrix related to $|\psi_{\in}\rangle$ is in the form
\begin{equation}
  \hat{\rho}_{\in}=
  \begin{pmatrix}
    \cos^{2}{\vartheta/2}&\frac{1}{2}e^{-i\varphi}\sin{\vartheta}\\
    \frac{1}{2}e^{i\varphi}\sin{\vartheta}&\sin^{2}{\vartheta/2}\,
  \end{pmatrix}.
\end{equation}
The output state $\hat{\rho}_{\out}$ can be obtained by applying a
joint measurement and local unitary transformation on the input
state $\hat{\rho}_{\in}$ \cite{Bowen}
\begin{equation}\label{P0Protocol}
\hat{\rho}_{\out}=\Lambda^{\hat{\rho}_{\ch},{\mathcal
P}_0}(\hat{\rho}_{\in})=\sum_{i=0}^{3}p_{i}\sigma^{i}\hat{\rho}_{\in}\sigma^{i},
\end{equation}
where $p_{i}=\Tr(E^i \hat{\rho}_{\ch})$ such that
$\sum_{i}p_{i}=1$. Here $E^i=|\Psi_{\Bell}^{i}\rangle\langle
\Psi_{\Bell}^{i}|$ where $|\Psi_{\Bell}^{i}\rangle$ are the four
maximally entangled Bell states associated with the Pauli matrices
$\sigma^i$, i.e. $E^i=(\sigma^i\otimes \sigma^0) E^0
(\sigma^i\otimes \sigma^0)$, where $\sigma^0=I$,
$\sigma^1=\sigma_x$, $\sigma^2=\sigma_y$ and $\sigma^3=\sigma_z$.
Furthermore for optimal utilization of a given entangled state as
resource, one must choose local basis states such that $p_0$ is
maximum, i.e. $p_0=\max\{p_i\}$. We therefore find that
$|\Psi_{\Bell}^{0}\rangle=\frac{1}{\sqrt{2}}\left(|+\rangle|v_{2}\rangle+|-\rangle|v_{1}\rangle\right)$,
$|\Psi_{\Bell}^{1}\rangle=\frac{1}{\sqrt{2}}\left(|+\rangle|v_{1}\rangle+|-\rangle|v_{2}\rangle\right)$,
$|\Psi_{\Bell}^{2}\rangle=\frac{1}{\sqrt{2}}\left(|+\rangle|v_{1}\rangle-|-\rangle|v_{2}\rangle\right)$,
$|\Psi_{\Bell}^{3}\rangle=\frac{1}{\sqrt{2}}\left(|+\rangle|v_{2}\rangle-|-\rangle|v_{1}\rangle\right)$,
and
\begin{eqnarray}
\nonumber p_0& = & \frac{1}{2}\left(1-x^2(t)\cos^{2}{\theta/2}
  +\frac{f(t)\sqrt{1-x^2(t)}}{x(t)}
  \sin{\theta}\cos{\phi}\right),\\
  p_{1}&=&p_{2}=\frac{1}{2}x^{2}(t)\cos^{2}{\theta/2},\\ \nonumber
  p_{3}&=&\frac{1}{2}\left(1-x^2(t)\cos^{2}{\theta/2}
  -\frac{f(t)\sqrt{1-x^2(t)}}{x(t)}
  \sin{\theta}\cos{\phi}\right).
 \end{eqnarray}
 Therefore according to equation (\ref{P0Protocol}), for the output we get
\begin{equation}
\nonumber
  \hat{\rho}_{\out}=
  \begin{pmatrix}
    \left(p_{0}+p_{3}\right)\cos^{2}{\vartheta/2}
    +2p_{1}\sin^{2}{\vartheta/2} &
    \frac{1}{2}\left(p_{0}-p_{3}\right)e^{-i\varphi}\sin{\vartheta}\\
    \frac{1}{2}\left(p_{0}-p_{3}\right)e^{i\varphi}\sin{\vartheta} &
    \left(p_{0}+p_{3}\right)\sin^{2}{\vartheta/2}
    +2p_{1}\cos^{2}{\vartheta/2}\
  \end{pmatrix}.
\end{equation}
To characterize the quality of the teleported state
 $\hat{\rho}_{\out}$, it is often quite useful to look at the fidelity
 between $\hat{\rho}_{\in}$ and $\hat{\rho}_{\out}$ defined as
 $F(\hat{\rho}_{\in},\hat{\rho}_{\out})=
\left[\textmd{Tr}\left(\sqrt{\sqrt{\hat{\rho}_{\in}}\hat{\rho}_{\out}\sqrt{\hat{\rho}_{\in}}}\right)\right]^2$,
\cite{uhlmann,jozsa}. For our system that the input state is pure,
the fidelity $F(\hat{\rho}_{\in},\hat{\rho}_{\out})$ can be easily
calculated as
\begin{equation}
F(\hat{\rho}_{\in},\hat{\rho}_{\out})=
\langle\psi_{\in}|\hat{\rho}_{\out}|\psi_{in}\rangle
 = (p_{0}+p_{3})+(p_1-p_3)\sin^{2}{\vartheta}.
\end{equation}
The average fidelity  is another useful concept for characterizing
the quality of teleportation and can be obtained by averaging the
fidelity $F(\hat{\rho}_{in},\hat{\rho}_{out})$ over all possible
input states
\begin{equation}
{\overline F}(\Lambda^{\hat{\rho}_{\ch},{\mathcal
P}_0})=\frac{1}{4\pi}\int_{0}^{2\pi}\d\varphi
\int_{0}^{\pi}F(\hat{\rho}_{\in},\hat{\rho}_{\out})\sin{\vartheta}\d\vartheta.
\end{equation}
For our system we get
\begin{equation}\label{FidelityA}
{\overline F}(\Lambda^{\hat{\rho}_{\ch},{\mathcal
P}_0})=\frac{2}{3}+\frac{1}{3}\left(\frac{f(t)\sqrt{1-x^2(t)}}{x(t)}
  \sin{\theta}\cos{\phi}-x^2(t)\cos^{2}{\theta/2}\right).
 \end{equation}
 We,  therefore, see that the atom-field entangled state $\hat{\rho}(t)$
 can be useful to transmit $|\psi_{\in}\rangle$ with fidelity better than any classical communication
 protocol, i.e. fidelity better that $2/3$, if we require that the second term in the above equation be strictly positive.

\begin{figure}[t]
\centerline{\includegraphics[height=6cm]{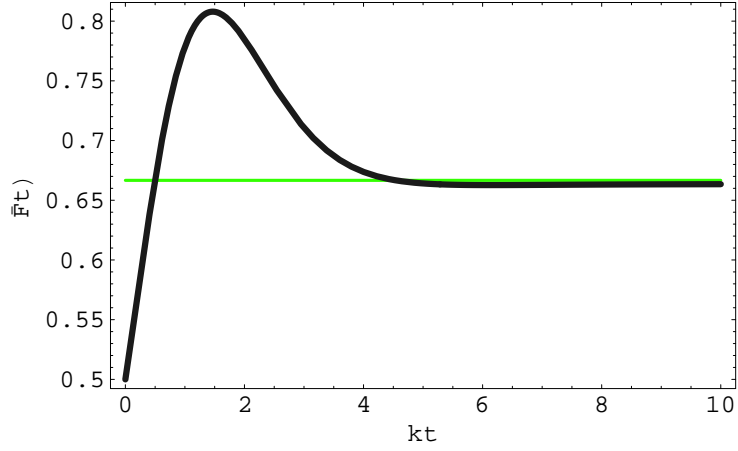}}
\caption{Optimal fidelity is plotted as a function of $kt$ with
$\theta=\pi/2$, $\phi=0$ and $g=1$. The horizontal line shows the
classical capacity $2/3$.}
\end{figure}
Horodecki et al have presented a beautiful formula relating the
optimal fidelity of teleportation and the maximal entangled fraction
\cite{Horodecki1999}. They have shown that for a given bipartite
state acting on ${\mathbb C}^d\otimes {\mathbb C}^d$, the optimal
fidelity of teleportation is given by
\begin{equation}\label{FidelityandFSF}
F_{\max}(\Lambda^{\hat{\rho}_{\ch},{\mathcal P}_0})=
\frac{f_{\max}(\Lambda^{\hat{\rho}_{\ch}})\;d+1}{d+1},
\end{equation}
where $f_{\max}(\Lambda^{\hat{\rho}_{\ch}})$ is the maximal
entangled fraction of the channel. Simple calculation shows that
in our model, i.e. $d=2$ and
$f_{\max}(\Lambda^{\hat{\rho}_{\ch}})=\max\left\{p_0,p_1,p_2,p_3\right\}=p_0$,
equation (\ref{FidelityandFSF}) gives the same result as equation
(\ref{FidelityA}). In Fig. 4 we plot the optimal fidelity of
teleportation as a function of $kt$ with $\theta=\pi/2$, $\phi=0$
and $g=1$. The horizontal line  shows the best classical fidelity
$2/3$.

\subsection{Entanglement teleportation}
We now consider the atom-field state as a quantum channel for
entanglement teleportation of a two-qubit state. We will consider
Lee and Kim's \cite{Lee2000} two-qubit teleportation protocol, and
use two copies of the above atom-field state as resource. In this
protocol, the joint measurement is decomposable into two
independent Bell measurements and the unitary operation into local
one-qubit Pauli rotations. Accordingly, for the output state we
get
 \begin{equation}
  \hat{\rho}_{\out}=\Lambda^{\hat{\rho}_{\ch},{\mathcal
P}_1}(\hat{\rho}_{\in})=\sum_{i,j=0}^{3}p_{ij}(\sigma_{i}\otimes\sigma_{j})\hat{\rho}_{\in}
  (\sigma_{i}\otimes\sigma_{j}),
\end{equation}
where
$p_{ij}=\Tr(E^{i}\hat{\rho}_{\ch})\Tr(E^{j}\hat{\rho}_{\ch})=p_ip_j$.
Here $E^{i}$ are projection  on the Bell states, defined in the
last section. We consider as input a two-qubit state in the
following pure state
$|\psi_{\in}\rangle=\cos{\vartheta/2}|+-\rangle+e^{i\varphi}\sin{\vartheta/2}|-+\rangle$
with $0\leq\vartheta\leq\pi,\;\;0\leq\varphi<2\pi$. The density
matrix related to $|\psi_{in}\rangle$ is in the form
\begin{equation}
\nonumber
  \hat{\rho}_{\in}=
  \begin{pmatrix}
    0 & 0 & 0 & 0 \\
    0 & a & c & 0 \\
    0 & c^{*} & b & 0 \\
    0 & 0 & 0 & 0\,
  \end{pmatrix},
\end{equation}
where we have defined $a=\cos^{2}{\vartheta/2}$,
$b=\sin^{2}{\vartheta/2}$ and
$c=\frac{1}{2}e^{-i\varphi}\sin{\vartheta}$. The concurrence of
this state is $C(\hat{\rho}_{\in})=\sin\vartheta$. For the output
we get
\begin{equation} \nonumber
  \hat{\rho}_{\out}=
  \begin{pmatrix}
   2p_{1}(p_{0}+p_{3}) & 0 & 0 & 0 \\
    0 & (p_{0}+p_{3})^{2}a+4p_{1}^{2}b & (p_{0}-p_{3})^{2}c & 0 \\
    0 & (p_{0}-p_{3})^{2}c^{*} & (p_{0}+p_{3})^{2}b+4p_{1}^{2}a  & 0 \\
    0 & 0 & 0 & 2p_{1}(p_{0}+p_{3}) \
  \end{pmatrix}.
\end{equation}
Now in order to calculate the concurrence of $\hat{\rho}_{\out}$,
we first calculate the eigenvalues of the operator
$\hat{\rho}_{\out}\tilde{\hat{\rho}}_{\out}$ as
\begin{eqnarray}
\nonumber
  \lambda_{1}&=& \left(\left(p_{0}+p_{3}\right)^{2}a+4p_{1}^{2}b\right)
  \left(\left(p_{0}+p_{3}\right)^{2}b+4p_{1}^{2}a\right)+
  \left(p_{0}-p_{3}\right)^{4}|c|^{2}\\ \nonumber
  &+&
  2\left(p_{0}-p_{3}\right)^{2}|c|\sqrt{\left(\left(p_{0}+p_{3}\right)^{2}a+4p_{1}^{2}b\right)
  \left(\left(p_{0}+p_{3}\right)^{2}b+4p_{1}^{2}a\right)},\\
  \lambda_{2} &=&
  \lambda_{3}=4\left(p_{0}+p_{3}\right)^{2}p_{1}^{2},\\\nonumber
 \lambda_{4}&=&
 \left(\left(p_{0}+p_{3}\right)^{2}a+4p_{1}^{2}b\right)
  \left(\left(p_{0}+p_{3}\right)^{2}b+4p_{1}^{2}a\right)+
  \left(p_{0}-p_{3}\right)^{4}|c|^{2}\\ \nonumber
  &-&
  2\left(p_{0}-p_{3}\right)^{2}|c|\sqrt{\left(\left(p_{0}+p_{3}\right)^{2}a+4p_{1}^{2}b\right)
  \left(\left(p_{0}+p_{3}\right)^{2}b+4p_{1}^{2}a\right)}.
\end{eqnarray}
Then by using equation (\ref{concurrence}), we obtain the
concurrence of the teleported  state $\hat{\rho}_{\out}$ as
\begin{eqnarray}\nonumber
C(\hat{\rho}_{\out})&=&\max\left\{0,\left(p_{0}-p_{3}\right)^{2}\sin\vartheta
-4\left(p_{0}+p_{3}\right)p_{1}\right\}\\
&=&\max\left\{0,\frac{(1-x^2)f^2}{x^2}\sin^2{\theta}\cos^2{\phi}\;\sin{\vartheta}-2x^2\left(1-x^2\cos^2{\theta/2}\right)
\cos^2{\theta/2}\right\}.
\end{eqnarray}

\begin{figure}[t]
\centerline{\includegraphics[height=10cm]{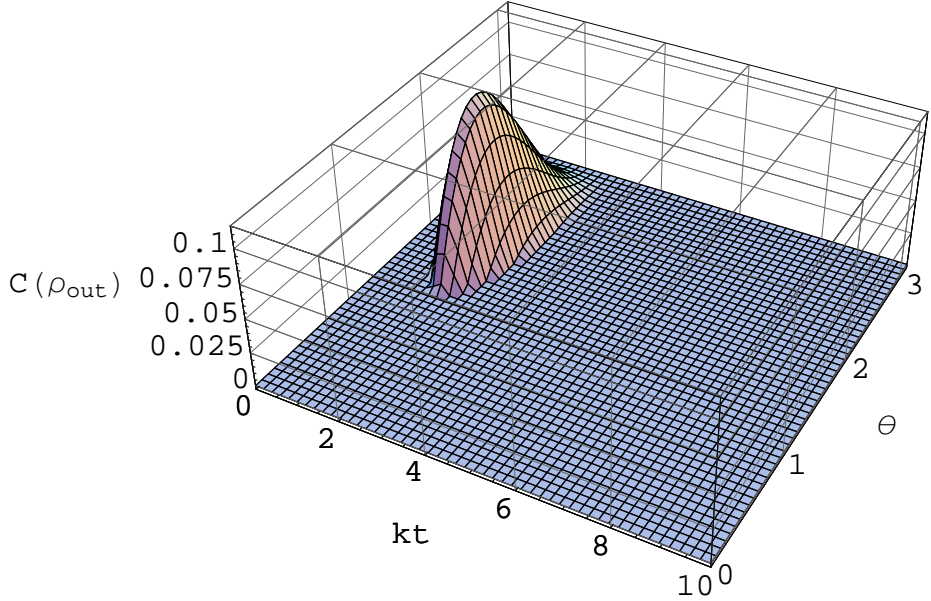}}
\caption{concurrence of the output state is plotted as a function of
$kt$ and $\theta$, with  $g=1$, $\phi=0$ and $\vartheta=\pi/2$.}
\end{figure}
In Fig. 5 we plot the concurrence of the output state as a
function of $kt$ and $\theta$. It is clear from the figure that
teleportation of a maximally Bell state (concurrence 1) via this
channel, give an output state with concurrence less than $0.1$. We
can also calculate the fidelity of $\hat{\rho}_{\in}$ and
$\hat{\rho}_{\out}$ as
\begin{equation}
F(\hat{\rho}_{\in},\hat{\rho}_{\out})=(p_{0}+p_{3})^{2}+2(
p_{1}^{2}-p_{0}p_{3})\sin^{2}\vartheta
\end{equation}
Now using equation (\ref{FidelityandFSF}) with $d=4$ and
$f_{\max}({\Lambda^{\hat{\rho}_{\ch}}})=p_0^2$,  the optimal
teleportation fidelity achievable is given by
\begin{eqnarray}\label{FidelityA3}\nonumber
{\overline F}({\Lambda^{\hat{\rho}_{\ch},{\mathcal
P}_1}})&=&\frac{2}{5}+\frac{1}{5}\left(\frac{f^2(1-x^2)}{x^2}\sin^2{\theta}\cos^2{\phi}
+x^4\cos^4{\theta/2}\right. \\
&+&\left.2(1-x^2\cos^2{\theta/2})\frac{f\sqrt{1-x^2}}{x}\sin{\theta}\cos{\phi}-2x^2\cos^2{\theta/2}\right).
\end{eqnarray}
In Fig. 6 we plot optimal fidelity of teleportation as a function
of $kt$ with $\theta=\pi/2$, $\phi=0$ and $g=1$. The horizontal
line  shows the classical fidelity $2/5$.
\begin{figure}[t]
\centerline{\includegraphics[height=5cm]{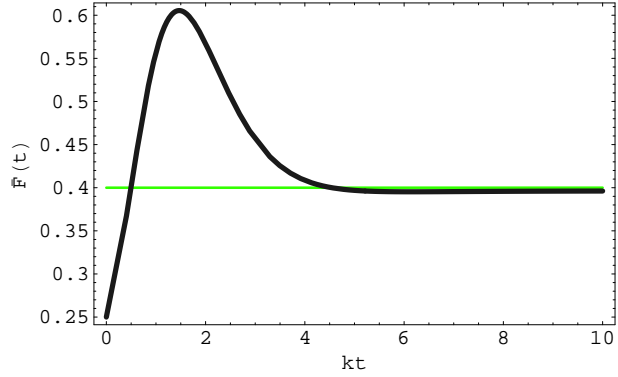}}
\caption{Optimal fidelity is plotted as a function of $kt$ with
$\theta=\pi/2$, $\phi=0$ and $g=1$. The horizontal line  shows the
classical fidelity $2/5$.}
\end{figure}

\section{Conclusion}
We have investigated the quantum entanglement and decoherence in the
interacting system of a two-level atom and a single mode vacuum
field in the presence of field dissipation. Starting from the cavity
field in a vacuum state and the atom in a general pure state, it is
shown that the final density matrix has support on ${\mathbb
C}^2\otimes{\mathbb C}^2$ space, i.e. the atom-field system
constitute a two-qubit system. We have, therefore, used the
concurrence as a relevant measure of entanglement between the atom
and the field. The effect of the atomic initial pure state on the
entanglement of the system is studied and it is shown that when the
atom is initially in the state
$|+\rangle=\frac{1}{\sqrt{2}}(|g\rangle+|e\rangle)$, the atom-field
entanglement is zero for all times. In this case we have shown that
the evolution of the system is unitary and therefore, the system
initial state $|+\rangle|0\rangle$ defines a decoherence-free
subspace. The influence of the cavity decay on the quantum
entanglement of the system has also been discussed and we have found
that the dissipation suppresses the entanglement. We have also
examined the Bell-CHSH violation between the atom and the field and
have shown that there are entangled states for which the Bell-BCSH
inequality is not violated. The decoherence induced by the cavity
decay is also studied and it is shown that the coherence properties
of the atom and also the field are affected by cavity decay.  The
one-qubit teleportation via the quantum channel constructed by the
atom-field system is also investigated. We have shown that the
atom-field entangled state can be useful to transmit a generic
one-qubit state $|\psi_{in}\rangle$ with fidelity better than any
classical communication protocol, i.e. fidelity better that $2/3$.
We have also studied the two-qubit entanglement teleportation via
two copies of the atom-field system. The fidelity of the
teleportation and also the entanglement of the replica are also
discussed and it is shown that the atom-field entangled state is
still superior to classical channel in performing the two-qubit
teleportation.

\newpage


\end{document}